\documentclass[twocolumn,a4paper,10pt,prl]{revtex4}%
\usepackage{amsmath}
\usepackage{graphicx}
\usepackage{amsfonts}
\usepackage{amssymb}%
\begin{document}
\preprint{quant-ph/ 0306126}
\title{Frequency up- and down-conversions in two-mode cavity quantum electrodynamics}
\author{R. M. Serra$^{1,2}$, C. J. Villas-B\^{o}as$^{1}$, N. G. de Almeida$^{3,4}$,
and M. H. Y. Moussa$^{1}$}
\affiliation{$^{1}$Departamento de F\'{\i}sica, Universidade Federal de S\~{a}o Carlos,
P.O. Box 676, S\~{a}o Carlos, 13565-905, S\~{a}o Paulo, Brazil\linebreak$^{2}%
$Optics Section, The Blackett Laboratory, Imperial College, London, SW7 2BZ,
United Kingdom\linebreak$^{3}$Departamento de Matem\'{a}tica e F\'{\i}sica,
Universidade Cat\'{o}lica de Goi\'{a}s, P.O. Box 86, Goi\^{a}nia,74605-010,
Goi\'{a}s, Brazil\linebreak$^{4}$Instituto de F\'{\i}sica, Universidade
Federal de Goi\'{a}s, Goi\^{a}nia, 74.001-970, GO, Brazil }

\begin{abstract}
In this letter we present a scheme for the implementation of frequency up- and
down-conversion operations in two-mode cavity quantum electrodynamics (QED).
This protocol for engineering bilinear two-mode interactions could enlarge
perspectives for quantum information manipulation and also be employed for
fundamental tests of quantum theory in cavity QED. As an application we show
how to generate a two-mode squeezed state in cavity QED (the original
entangled state of Einstein-Podolsky-Rosen).


\end{abstract}
\maketitle

Parametric frequency conversion has been a major ingredient in quantum optics.
Employed in the generation of squeezed and two-photon states of light to test
sub-poissonian statistics \cite{Stoler} and Bell's inequalities \cite{Kwiat},
parametric down-conversion (PDC) has been constantly revisited through the
work by Louisell \textit{et al.} \cite{Louisell}. Sub-poissonian statistics,
one of the characteristics of squeezed light, has deepened our understanding
of the properties of radiation \cite{Stoler} and its interaction with matter
\cite{Milburn}. It has provided an unequivocal signature of the quantum nature
of light, disputed since the discovery of the photoelectric effect, and has
continued to motivate fundamental works up to the present \cite{Haroche}.
Apart from fundamental phenomena, the potential application of PDC in
technology is also striking, ranging from improvements in the signal to noise
ratio in optical communication \cite{SN} to the possibility of measuring
gravitational waves through squeezed fields \cite{Caves}.

The combination of simplicity and comprehensiveness exhibited by the
frequency-conversion mechanisms applied in some of the recent proposals of
quantum information theory \cite{Pittman}\textbf{ }has motivated the goal of
the present letter: the implementation of the frequency up- and
down-conversion operations in two-mode cavity quantum electrodynamics (QED).
With this protocol to engineer two-mode interactions, it would be possible to
map into cavity QED some of the proposals for quantum logical processing
originally designed for travelling fields. This protocol may be useful for
scalable quantum computation and communication proposals \cite{CZ}, besides
enlarging such perspectives, it may also be employed for fundamental tests of
quantum theory \cite{Rausch}.\textbf{ }

The parametric frequency conversion operations are accomplished through the
dispersive interactions of the cavity modes with a single three-level-driven
atom injected into the cavity, which works as a nonlinear medium. Although
considerable space has been devoted in the literature to the interaction
between a three-level atom and two cavity modes \cite{Walls}, the issue of
tailoring the bilinear Hamiltonians of frequency conversion processes in
cavity QED has not been addressed.

\textit{Parametric up conversion (PUC).} We envisage working with Rydberg
atoms in the microwave regime. Starting with the PUC, the energy diagram of
the Rydberg three-level atom is in the $\Lambda$ configuration as sketched in
Fig. 1a. The ground ($\left\vert g\right\rangle $) and excited ($\left\vert
e\right\rangle $) states are coupled through an auxiliary more-excited level
($\left\vert i\right\rangle $). The cavity microwave modes of frequencies
$\omega_{a}$ and $\omega_{b}$ enable both dipole-allowed transitions
$\left\vert g\right\rangle $ $\leftrightarrow$ $\left\vert i\right\rangle $
and $\left\vert e\right\rangle $ $\leftrightarrow$ $\left\vert i\right\rangle
$, with coupling constants $\lambda_{a}$ and $\lambda_{b}$, respectively, and
detuning $\Delta$ $=\omega_{i}-\omega_{g}-\omega_{a}=\omega_{i}-\omega
_{e}-\omega_{b}$. Finally, a classical field of frequency $\omega_{0}%
=\omega_{e}-\omega_{g}-\delta$, dispersively driving the dipole-forbidden
atomic transition $\left\vert g\right\rangle $ $\leftrightarrow$ $\left\vert
e\right\rangle $ with coupling constant $\Omega$, leads to the desired
interaction between the modes $\omega_{a}$ and $\omega_{b}$.{\Large \ }This
dipole-forbidden transition can be induced by applying a sufficiently strong
electric field. The Hamiltonian which describe this system, in the interaction
picture within the rotating wave approximation and in a rotating frame
(through the unitary transformation $\exp\left[  -i\Delta t\left(  \sigma
_{ee}+\sigma_{gg}\right)  \right]  $), is given by
\begin{align}
\mathbf{H}  &  =\hbar\left(  \lambda_{a}a\sigma_{ig}+\lambda_{b}b\sigma
_{ie}+\Omega\operatorname*{e}\nolimits^{-i\delta t}\sigma_{ge}+\mathrm{{h.c.}%
}\right) \nonumber\\
&  -\hbar\Delta\left(  \sigma_{ee}+\sigma_{gg}\right)  . \label{Eq2}%
\end{align}
with $a^{\dagger}$ ($a$) and $b^{\dagger}$ ($b$) standing for the creation
(annihilation) operators of the quantized cavity modes, while $\sigma
_{kl}\equiv\left\vert k\right\rangle \left\langle l\right\vert $ $(k,l=g,e,i)$
defines the atomic transition operators.

Considering the Heisenberg equations of motion for the transition operators
$\sigma_{ig}$ and $\sigma_{ei}$, we can compare the time scales of the
transitions induced by the cavity modes. If the dispersive transitions induced
by the quantized fields are sufficiently{\large \ }detuned, i.e., $\Delta\gg$
$\left\vert \lambda_{a}\right\vert $,$\left\vert \lambda_{b}\right\vert
$,$\left\vert \Omega\right\vert $, we obtain the adiabatic solutions for the
transition operators $\sigma_{ig}$ and $\sigma_{ie}$ by setting $d\sigma
_{ig}/dt=d\sigma_{ie}/dt=0$, given by
\begin{subequations}
\begin{align}
\sigma_{ig}  &  \simeq\left(  \frac{\Delta^{2}}{\Delta^{2}+\Omega^{2}}\right)
\left[  \frac{\lambda_{a}^{\ast}}{\Delta}a^{\dagger}\left(  \sigma_{ii}%
-\sigma_{gg}\right)  -\frac{\lambda_{b}^{\ast}}{\Delta}b^{\dagger}\sigma
_{eg}\right. \nonumber\\
&  \left.  -\frac{\Omega\lambda_{a}^{\ast}}{\Delta^{2}}\operatorname*{e}%
\nolimits^{-i\delta t}a^{\dagger}\sigma_{ge}+\frac{\Omega\lambda_{b}^{\ast}%
}{\Delta^{2}}\operatorname*{e}\nolimits^{-i\delta t}b^{\dagger}\left(
\sigma_{ii}-\sigma_{ee}\right)  \right]  , \label{Eq4a}%
\end{align}%
\end{subequations}
\begin{subequations}
\begin{align}
\sigma_{ie}  &  \simeq\left(  \frac{\Delta^{2}}{\Delta^{2}+\Omega^{2}}\right)
\left[  -\frac{\lambda_{a}^{\ast}}{\Delta}a^{\dagger}\sigma_{ge}+\frac
{\lambda_{b}^{\ast}}{\Delta}b^{\dagger}\left(  \sigma_{ii}-\sigma_{ee}\right)
\right. \nonumber\\
&  \left.  +\frac{\Omega\lambda_{a}^{\ast}}{\Delta^{2}}\operatorname*{e}%
\nolimits^{i\delta t}a^{\dagger}\left(  \sigma_{ii}-\sigma_{gg}\right)
-\frac{\Omega\lambda_{b}^{\ast}}{\Delta^{2}}\operatorname*{e}%
\nolimits^{i\delta t}b^{\dagger}\sigma_{eg}\right]  . \label{Eq4bb}%
\end{align}
Inserting these adiabatic solutions for $\sigma_{ig}$ and $\sigma_{ie}$ into
Eq. (\ref{Eq2}), the following Hamiltonian is obtained (assuming $\Delta
^{2}+\Omega^{2}\approx\Delta^{2}$)
\end{subequations}
\begin{align}
\mathbf{H}  &  \simeq-\hbar\left(  \Delta+\frac{\left\vert \lambda
_{b}\right\vert ^{2}}{\Delta}\right)  \sigma_{ee}-\hbar\left(  \Delta
+\frac{\left\vert \lambda_{a}\right\vert ^{2}}{\Delta}\right)  \sigma
_{gg}\nonumber\\
&  +\hbar\frac{\left(  \left\vert \lambda_{a}\right\vert ^{2}+\left\vert
\lambda_{b}\right\vert ^{2}\right)  }{\Delta}\sigma_{ii}\nonumber\\
&  +\hbar\Omega\left(  1-\frac{\left(  \left\vert \lambda_{a}\right\vert
^{2}+\left\vert \lambda_{b}\right\vert ^{2}\right)  }{2\Delta^{2}}\right)
\left(  \operatorname*{e}\nolimits^{-i\delta t}\sigma_{ge}+\mathrm{{h.c.}%
}\right) \nonumber\\
&  +\frac{\hbar}{\Delta}\left.  \left(  \left\vert \lambda_{a}\right\vert
^{2}a^{\dagger}a+\left\vert \lambda_{b}\right\vert ^{2}b^{\dagger}b\right)
\sigma_{ii}\right] \nonumber\\
&  \left.  -\left\vert \lambda_{a}\right\vert ^{2}a^{\dagger}a\sigma
_{gg}-\left\vert \lambda_{b}\right\vert ^{2}b^{\dagger}b\sigma_{ee}\right]
\nonumber\\
&  -\frac{\hbar\Omega}{\Delta^{2}}\left(  \left\vert \lambda_{a}\right\vert
^{2}a^{\dagger}a+\left\vert \lambda_{b}\right\vert ^{2}b^{\dagger}b\right)
\left(  \operatorname*{e}\nolimits^{-i\delta t}\sigma_{ge}+\mathrm{{h.c.}%
}\right) \nonumber\\
&  -\frac{\hbar}{\Delta}\left(  \lambda_{a}\lambda_{b}^{\ast}ab^{\dagger
}\sigma_{eg}+\mathrm{{h.c.}}\right) \nonumber\\
&  +\frac{\hbar\Omega}{\Delta^{2}}\left(  \lambda_{a}\lambda_{b}^{\ast
}\operatorname*{e}\nolimits^{-i\delta t}ab^{\dagger}+\mathrm{{h.c.}}\right)
\left(  \sigma_{ii}-\sigma_{gg}-\sigma_{ee}\right)  . \label{Eq5}%
\end{align}%
%TCIMACRO{\FRAME{fthFU}{3.0165in}{4.3353in}{0pt}{\Qcb{Energy diagram of a
%three-level atom in the (a) $\Lambda$ configuration to obtain the PUC process
%and in the (b) ladder configuration to obtain the PDC process.}}%
%{}{fig1ab-all.ps}{\special{ language "Scientific Word";  type "GRAPHIC";
%maintain-aspect-ratio TRUE;  display "USEDEF";  valid_file "F";
%width 3.0165in;  height 4.3353in;  depth 0pt;  original-width 7.8646in;
%original-height 11.3334in;  cropleft "0";  croptop "1";  cropright "1";
%cropbottom "0";  filename 'fig1ab-all.ps';file-properties "XNPEU";}}}%
%BeginExpansion
\begin{figure}
[th]
\begin{center}
\includegraphics[
height=4.3353in,
width=3.0165in
]%
{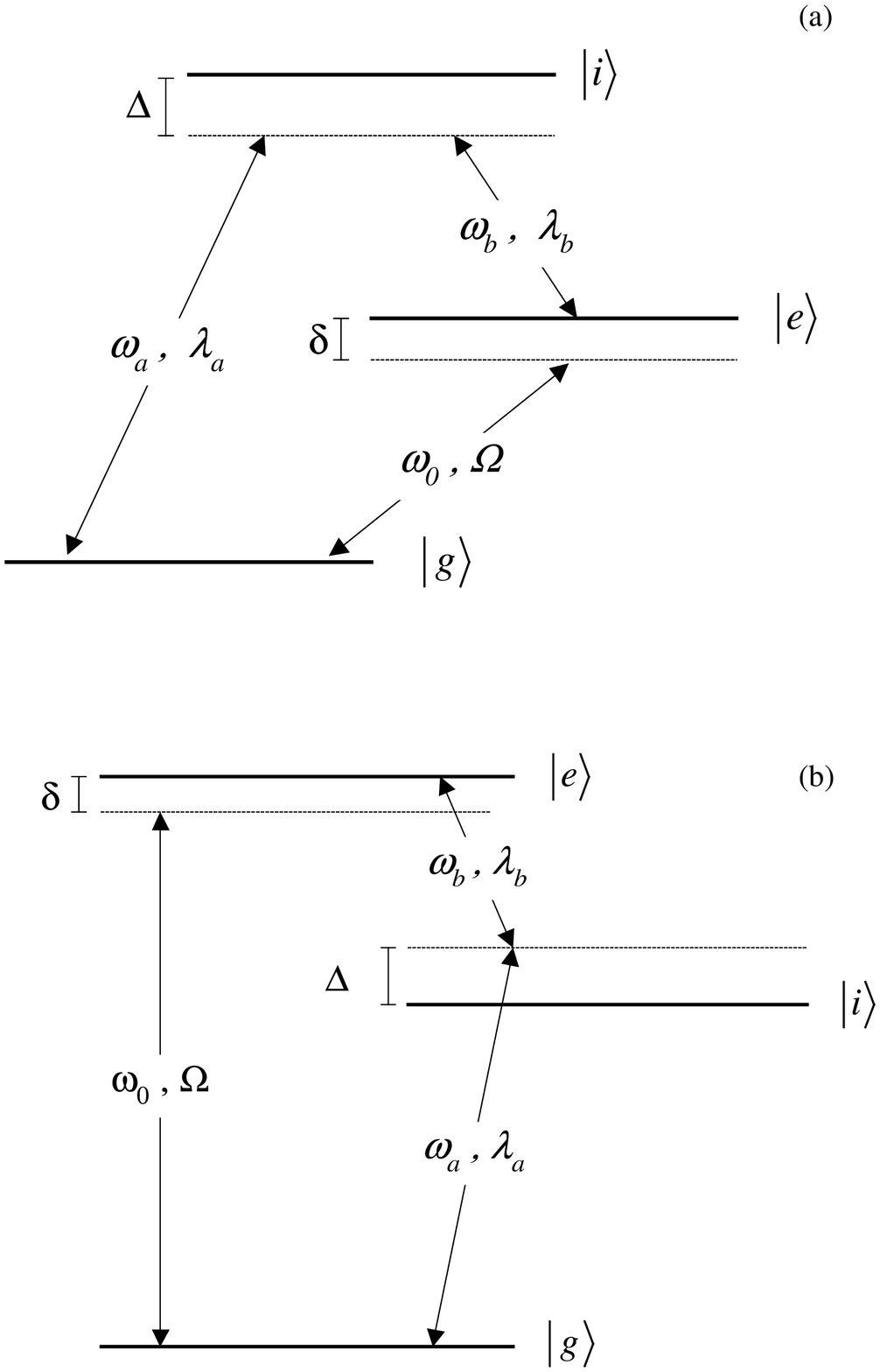}%
\caption{Energy diagram of a three-level atom in the (a) $\Lambda$
configuration to obtain the PUC process and in the (b) ladder configuration to
obtain the PDC process.}%
\end{center}
\end{figure}
%EndExpansion

The state vector associated with Hamiltonian (\ref{Eq5}), in the interaction
picture, can be written as $|\Psi\left(  t\right)  \rangle=\sum_{\ell
=g,e,i}\left\vert \ell\right\rangle \left\vert \Phi_{\ell}\left(  t\right)
\right\rangle \mathrm{{,}}$ where $|\Phi_{\ell}\left(  t\right)
\rangle=\widehat{1}_{ab}\otimes\left\langle \ell\right.  |\Psi\left(
t\right)  \rangle$ and $\widehat{1}_{ab}$ is the unitary operator of cavity
modes represented in a convenient basis. Using the orthogonality of the atomic
states in $|\Psi\left(  t\right)  \rangle$ and Eq. (\ref{Eq5}) we obtain the
uncoupled time-dependent (TD) Schr\"{o}dinger equations for the atomic
subspace $\left\vert i\right\rangle $ (in the interaction picture),
$i\hbar\frac{d}{dt}|\Phi_{i}\left(  t\right)  \rangle=\mathcal{H}_{i}|\Phi
_{i}\left(  t\right)  \rangle\mathrm{{,}}$ with $\mathcal{H}_{i}=\hbar\chi
_{a}a^{\dagger}a+\hbar\chi_{b}b^{\dagger}b+\hbar\left(  \xi\operatorname*{e}%
\nolimits^{-i\delta t}ab^{\dagger}+\xi^{\ast}\operatorname*{e}%
\nolimits^{i\delta t}a^{\dagger}b\right)  \mathrm{{,}}$where $\chi_{\ell
}=\left\vert \lambda_{\ell}\right\vert ^{2}/\Delta$ ($\ell=a,b$) stand for the
shift factors of the two cavity-mode frequencies, while $\xi=\Omega\lambda
_{a}\lambda_{b}^{\ast}/\Delta^{2}$ is the effective coupling parameter between
these modes. The subscript $i$ indicates the atomic subspace $\left\vert
i\right\rangle $. The TD Schr\"{o}dinger equations for subspace $\left\{
\left\vert g\right\rangle ,\left\vert e\right\rangle \right\}  $, which follow
from Eq. (\ref{Eq5}), couple the fundamental and excited atomic states.
Therefore, when we prepare the initial state of the atom in the auxiliary
level $\left\vert i\right\rangle $, the dynamics of the atom-field dispersive
interactions, governed by the effective Hamiltonian $\mathcal{H}_{i}$, results
in cavity modes with shifted frequencies which are coupled in identical
fashion to running waves crossing a nonlinear crystal, as in PUC.

Performing a unitary transformation on the Schr\"{o}dinger equation for
$|\Phi_{i}\left(  t\right)  \rangle$, through the operator $\exp\left[
-it\left(  \chi_{a}a^{\dagger}a+\chi_{b}b^{\dagger}b\right)  \right]  $, we
obtain the Hamiltonian $\widetilde{\mathcal{H}}_{i}=\hbar\left(  \xi
ab^{\dagger}\operatorname*{e}\nolimits^{-i\left(  \delta+\chi_{a}-\chi
_{b}\right)  t}+\xi^{\ast}a^{\dagger}b\operatorname*{e}\nolimits^{i\left(
\delta+\chi_{a}-\chi_{b}\right)  t}\right)  $. At this point we observe that
the choice of $\delta=\left(  \left\vert \lambda_{b}\right\vert ^{2}%
-\left\vert \lambda_{a}\right\vert ^{2}\right)  /\Delta$, the detuning
associated with the classical driving field, leads to the simplified form
\begin{equation}
\widetilde{\mathcal{H}}_{i}=\hbar\left(  \xi ab^{\dagger}+\xi^{\ast}%
a^{\dagger}b\right)  , \label{Eq10a}%
\end{equation}
where the up-conversion process for the effective frequencies is such that
$\omega_{0}=\left(  \omega_{a}+\chi_{a}\right)  -\left(  \omega_{b}+\chi
_{b}\right)  $. It should be noted that the degenerate up-conversion process
is the equivalent of a beam-splitter operation, which has been generally
required for quantum logical purposes \cite{Pittman}.

\textit{Parametric down-conversion (PDC).} Next, to engineer the PDC process,
we consider the atomic levels in the ladder configuration, as shown in Fig.
1(b), where the ground and excited states are coupled through an intermediate
level. The cavity microwave modes $\omega_{a}$ and $\omega_{b}$ are tuned to
the vicinity of the dipole-allowed transitions $\left\vert g\right\rangle $
$\leftrightarrow$ $\left\vert i\right\rangle $ and $\left\vert e\right\rangle
$ $\leftrightarrow$ $\left\vert i\right\rangle $ with coupling constants
$\lambda_{a}$ and $\lambda_{b}$, respectively, and detuning $\Delta$
$=-\left(  \omega_{i}-\omega_{g}-\omega_{a}\right)  =\omega_{e}-\omega
_{i}-\omega_{b}$. The desired interaction between the modes $\omega_{a}$\ and
$\omega_{b}$\ is accomplished by dispersively driving \ the dipole-forbidden
atomic transition $\left\vert g\right\rangle $ $\leftrightarrow$ $\left\vert
e\right\rangle $ with a classical field of frequency $\omega_{0}=\omega
_{e}-\omega_{g}-\delta$ and coupling constant $\Omega$.{\Large \ }The
Hamiltonian to engineer the PDC, within the rotating wave approximation, is
given by $H=H_{0}+V$, where
\begin{subequations}
\begin{align}
H_{0}  &  =\hbar\omega_{a}a^{\dagger}a+\hbar\omega_{b}b^{\dagger}b+\sum
_{\ell=g,e,i}\hbar\omega_{\ell}\sigma_{\ell\ell},\label{Eq11a}\\
V  &  =\hbar\left(  \lambda_{a}a\sigma_{ig}+\lambda_{b}b\sigma_{ei}+\Omega
e^{i\omega_{0}t}\sigma_{ge}+\mathrm{{h.c.}}\right)  {.} \label{Eq11b}%
\end{align}
Applying the transformation\ $\exp\left[  i\Delta t\left(  \sigma_{ee}%
+\sigma_{gg}\right)  \right]  $ to $H$ and following the steps leading from
Eq. (\ref{Eq2}) to (\ref{Eq5}) (considering as in PUC case $\Delta\gg$
$\left\vert \lambda_{a}\right\vert $,$\left\vert \lambda_{b}\right\vert
$,$\left\vert \Omega\right\vert $), we obtain the effective Hamiltonian (in
the interaction picture)
\end{subequations}
\begin{align}
\mathbf{H}  &  \simeq\hbar\left(  \Delta+\frac{\left\vert \lambda
_{a}\right\vert ^{2}}{\Delta^{2}}\right)  \sigma_{gg}+\hbar\left(
\Delta+\frac{\left\vert \lambda_{b}\right\vert ^{2}}{\Delta}\right)
\sigma_{ee}\nonumber\\
&  -\frac{\hbar\left(  \left\vert \lambda_{a}\right\vert ^{2}+\left\vert
\lambda_{b}\right\vert ^{2}\right)  }{\Delta}\sigma_{ii}+\frac{\hbar}{\Delta
}\left(  \lambda_{a}\lambda_{b}ab\sigma_{eg}+\mathrm{{h.c.}}\right)
\nonumber\\
&  +\hbar\Omega\left(  1-\frac{\left\vert \lambda_{a}\right\vert
^{2}+\left\vert \lambda_{b}\right\vert ^{2}}{2\Delta^{2}}\right)  \left(
e^{-i\delta t}\sigma_{ge}+\mathrm{{h.c.}}\right) \nonumber\\
&  +\frac{\hbar}{\Delta}\left[  -\left(  \left\vert \lambda_{a}\right\vert
^{2}a^{\dagger}a+\left\vert \lambda_{b}\right\vert ^{2}b^{\dagger}b\right)
\sigma_{ii}+\left\vert \lambda_{a}\right\vert ^{2}a^{\dagger}a\sigma
_{gg}+\left\vert \lambda_{b}\right\vert ^{2}b^{\dagger}b\sigma_{ee}\right] \\
&  -\frac{\hbar\Omega}{\Delta^{2}}\left(  \left\vert \lambda_{a}\right\vert
^{2}a^{\dagger}a+\left\vert \lambda_{b}\right\vert ^{2}b^{\dagger}b\right)
\left(  e^{-i\delta t}\sigma_{ge}+\mathrm{{h.c.}}\right) \nonumber\\
&  -\frac{\hbar\Omega}{\Delta^{2}}\left(  \lambda_{a}\lambda_{b}e^{-i\delta
t}ab+\mathrm{{h.c.}}\right)  \left(  \sigma_{ee}-\sigma_{ii}+\sigma
_{gg}\right)  . \label{Eq12}%
\end{align}

Next, expanding the state vector of the system as in the PUC case and
preparing the initial state of the atom in the auxiliary level, we obtain the
uncoupled time-dependent (TD) Schr\"{o}dinger equations for the atomic
subspace $\left\vert i\right\rangle $, with $\mathcal{H}_{i}=-\hbar\chi
_{a}a^{\dagger}a-\hbar\chi_{b}b^{\dagger}b+\hbar\left(  \xi e^{-i\delta
t}ab+\xi^{\ast}e^{i\delta t}a^{\dagger}b^{\dagger}\right)  $, where
$\xi=\Omega\lambda_{a}\lambda_{b}/\Delta^{2}$ is the effective coupling
parameter between these modes. Therefore, when we prepare the initial state of
the atom in the auxiliary level, the dynamics of the atom-field dispersive
interactions leads to shifted cavity modes which are coupled in identical
fashion to PDC. Performing the unitary transformation, $\exp\left[  it\left(
\chi_{a}a^{\dagger}a+\chi_{b}b^{\dagger}b\right)  \right]  $, we obtain the
Hamiltonian $\widetilde{\mathcal{H}}_{i}=\hbar\left(  \xi e^{-i\left(
\delta-\chi_{a}-\chi_{b}\right)  t}ab+\xi^{\ast}e^{i\left(  \delta-\chi
_{a}-\chi_{b}\right)  t}a^{\dagger}b^{\dagger}\right)  .$The choice
$\delta=\left(  \left\vert \lambda_{a}\right\vert ^{2}+\left\vert \lambda
_{b}\right\vert ^{2}\right)  /\Delta$ leads to the simplified form (where the
down-conversion process for the effective frequencies satisfies $\omega
_{0}=\left(  \omega_{a}-\chi_{a}\right)  +\left(  \omega_{b}-\chi_{b}\right)
$)
\begin{equation}
\widetilde{\mathcal{H}}_{i}=\hbar\left(  \xi ab+\xi^{\ast}a^{\dagger
}b^{\dagger}\right)  \mathrm{{.}} \label{Eq16}%
\end{equation}

\textit{Two-photon processes}. We obtain from Hamiltonians (\ref{Eq5}) and
(\ref{Eq12}), by switching off the classical amplification process (apart from
diagonal terms), the interactions $\hbar\left(  \zeta ab^{\dagger}\sigma
_{eg}+\zeta^{\ast}a^{\dagger}b\sigma_{ge}\right)  $ \cite{Gerry} and
$\hbar\left(  \kappa ab\sigma_{eg}+\kappa^{\ast}a^{\dagger}b^{\dagger}%
\sigma_{ge}\right)  $ \cite{Duvida}, respectively. The coupling parameters
read $\zeta=\hbar\lambda_{a}\lambda_{b}^{\ast}/\Delta$ and{\large \ }%
$\kappa=\hbar\lambda_{a}\lambda_{b}/\Delta${\large .} With these interactions
it is straightforward to prepare the Bell basis states for the cavity modes
($\left\vert \Psi_{ab}^{\pm}\right\rangle =\left(  \left\vert 1_{a}%
0_{b}\right\rangle \pm\left\vert 0_{a}1_{b}\right\rangle \right)  /\sqrt{2}$,
$\left\vert \Phi_{ab}^{\pm}\right\rangle =\left(  \left\vert 1_{a}%
1_{b}\right\rangle \pm\left\vert 0_{a}0_{b}\right\rangle \right)  /\sqrt{2}$),
with the passage of a single atom through the cavity. Moreover, as a
by-product of the present scheme, in the case where $\omega_{a}=\omega_{b}$
(discussed in detail in Ref. \cite{Celso}) we get the degenerate PDC process
corresponding to the well-known interaction $\hbar\left[  \xi\left(  a\right)
^{2}+\xi^{\ast}\left(  a^{\dagger}\right)  ^{2}\right]  $ which has been used
to generate squeezed states of light in cavity QED. We emphasize that \ this
degenerate down-conversion process can be used to squeeze an arbitrary state
previously prepared in the cavity; i.e., to perform the operation $S\left\vert
\Psi\right\rangle $ in cavity QED ($S$ being the squeeze operator)
\cite{Celso}.

\textit{The original EPR state. }As another application of the present
proposal, we derive the original EPR state expanded in the position
representation. Starting with the two cavity modes in their vacuum states and
applying the down-conversion interaction Eq. (\ref{Eq16}) during the time
interval $\tau$, following the procedure described above, the evolved two-mode
state reads (in the interaction picture)
\begin{align}
\left\vert \psi(\tau)\right\rangle _{ab}  &  =\operatorname*{e}%
\nolimits^{-i\tau\left(  \xi ab+\xi^{\ast}a^{\dagger}b^{\dagger}\right)
}\left\vert 0,0\right\rangle _{ab}\label{Eq17}\\
&  =\sum_{n=0}^{\infty}\frac{\left[  \tanh\left(  \left\vert \xi\right\vert
\tau\right)  \right]  ^{n}}{\cosh\left(  \left\vert \xi\right\vert
\tau\right)  }\left\vert n,n\right\rangle _{ab}\mathrm{{,}}%
\end{align}
where we have adjusted the coupling constants $\lambda_{a}$ and $\lambda_{b}$
such that $\xi=i\left\vert \xi\right\vert $. This state is the two-mode
squeezed vacuum state which, in the limit $\left\vert \xi\right\vert
\tau\rightarrow\infty$ (and projected into the positional basis of modes $a$
and $b$), is exactly the original entanglement used in the EPR argument
against the uncertainty principle \cite{EPR}. In order to estimate the
\textquotedblleft quality\textquotedblright\ of the prepared EPR state
(\ref{Eq17}), we compute, in this state, the mean values \cite{SB} $\left(
\Delta x\right)  ^{2}=\left\langle \left(  x_{a}-x_{b}\right)  ^{2}%
\right\rangle =\left.  \operatorname*{e}^{-2\left\vert \xi\right\vert \tau
}\right/  2$ and $(\Delta p)^{2}=\left\langle \left(  p_{a}+p_{b}\right)
^{2}\right\rangle =\left.  \operatorname*{e}^{-2\left\vert \xi\right\vert
\tau}\right/  2$, where $x_{\beta}=\left(  \beta+\beta^{\dagger}\right)  /2$
and $p_{\beta}=-i(\beta-\beta^{\dagger})/2$ ($\beta=a,b$) are the field
quadratures. We obtain\ the result $\left(  \Delta x\right)  ^{2}+(\Delta
p)^{2}=\operatorname*{e}\nolimits^{-2\left\vert \xi\right\vert \tau}$ which
goes to zero for the ideal EPR state ($\left\vert \xi\right\vert
\tau\rightarrow\infty$) and to unity for an entirely separable state
\cite{SB}. Therefore, the expression $1-\operatorname*{e}%
\nolimits^{-2\left\vert \xi\right\vert \tau}$ can be used to estimate the
quality of the prepared state (\ref{Eq17}) with present-day cavity QED
parameters. For specific cavity modes and atomic system, the interaction
parameter $\left\vert \xi\right\vert \tau$ can be adjusted in accordance with
the coupling strength $\xi$ (calculated from the parameters $\Omega$,
$\lambda_{a}$, $\lambda_{b}$, and $\Delta$) and the interaction time $\tau$.
Assuming typical values for the parameters involved, arising from Rydberg
states where the intermediate state $\left\vert i\right\rangle $, an
$(n-1)P_{3/2}$ level, is nearly halfway between $\left\vert g\right\rangle $,
an $(n-1)S_{1/2}$ level, and $\left\vert e\right\rangle $, an $nS_{1/2}$
level, we get $\left\vert \lambda_{a}\right\vert \sim\left\vert \lambda
_{b}\right\vert \sim7\times10^{5}$s$^{-1}$ \cite{BRH}. With these values and
assuming the detuning $\Delta\sim10^{7}$s$^{-1}$ (note that $\Delta
\sim14\lambda_{a,b}$ \cite{Allen}) and the coupling strength $\Omega
\sim7\times10^{5}$s$^{-1}$, we obtain $\left\vert \xi\right\vert \sim
3.4\times10^{3}$s$^{-1}$. For an atom-field interaction time about $\tau
\sim2\times10^{-4}$s, we get the interaction parameter $\left\vert
\xi\right\vert \tau\sim0.68$, close to the value ($0.69$) achieved for
building the EPR state for unconditional quantum teleportation in the
running-wave domain \cite{Furusawa}. The value $\left\vert \xi\right\vert
\tau\sim0.68$ leads to $1-\operatorname*{e}\nolimits^{-2\left\vert
\xi\right\vert \tau}$ $\sim0.74$, and we note that increasing moderately the
interaction time to $\tau\sim6\times10^{-4}$s ($\left\vert \xi\right\vert
\tau\sim2$) the quality of the prepared state increases to
$1-\operatorname*{e}\nolimits^{-2\left\vert \xi\right\vert \tau}$ $\sim0.98$.
Regarding the degenerate PDC process ($\omega_{a}=\omega_{b}$) \cite{Celso},
for an atom-field interaction time about $\tau\sim2\times10^{-4}$s we get the
squeezing factor $r=2\left\vert \xi\right\vert \tau\sim1.36$, such that the
variance in the squeezed quadrature turns to be $\operatorname{e}^{-2r}%
/4\sim1.6\times10^{-2}$, representing a squeezing up to 93\% (for an initial
coherent state prepared in the cavity) with the passage of just one atom.

There are some sensitive points in the experimental implementation of the
present scheme. The atomic detection efficiency and the spread of the atomic
velocity do not play important roles in the present scheme were only one step
of atom-fields interactions is required. However, due to the Gaussian profile
$f(x)$ of the cavity fields in the transverse direction, the atom-field
couplings $\lambda_{a}$ and $\lambda_{b}$ become time-dependent parameters as
well as the effective coupling between the cavity modes $\xi=\Omega\lambda
_{a}\lambda_{b}\left[  f(x)\right]  ^{2}/\Delta^{2}$ (where $f(x)=\exp
(-x^{2}/w^{2})$, $x$ is the time-dependent atom position from the center of
the cavity, and $w\sim0.6$ cm \cite{Haroche01} is the waist of the Gaussian).
The effect of the field profile can be evaluated straightforward by using the
analytical results for a time-dependent degenerate PDC process, demonstrated
in \cite{Celso01}, leading to the squeezing factor $r=\left(  \left.
\Omega\lambda_{a}\lambda_{b}\right/  \Delta^{2}\right)  2\int_{0}^{\tau
}\left[  f(x)\right]  ^{2}dt$. Considering the atom-field interaction time
about $\tau\sim2\times10^{-4}$s, we get the squeezing factor $r\sim0.51$. To
obtain the same value $r\sim1.36$ of the ideal case, we must increase
moderately the interaction time to $\tau\sim5.32\times10^{-4}$s. The
interaction times cited above are at least one order of magnitude smaller than
the decay time of the open cavities used in cavity QED experiments
\cite{Rausch,Haroche01}. Regarding atomic decay, we note that for Rydberg
levels the damping effects can be safely neglected for typical interaction
time scales.

We note that, to characterize the entangled state in (\ref{Eq17}) we can use
the reconstruction technique presented in \cite{FLD}. To employ this technique
we have firstly to apply the displacement operator $D^{-1}(\eta_{\ell}%
,\eta_{\ell}^{\ast})=\exp(-\eta_{\ell}\ell^{\dagger}+\eta_{\ell}^{\ast}\ell)$,
with $\ell=a,b$, into the cavity modes. Next, an additional three-level atom
is sent through the cavity, prepared in the superposition state $\left(
\left\vert i\right\rangle +\left\vert f\right\rangle \right)  /\sqrt{2}$,
where $\left\vert f\right\rangle $ stands for an auxiliary Rydberg level whose
transitions to the states $\left\vert g\right\rangle $, $\left\vert
i\right\rangle $, and $\left\vert e\right\rangle $ do not couple to the cavity
modes. Turning off the classical amplification field and considering the
atom-fields interaction time $t$, we obtain from Hamiltonian (\ref{Eq12}), the
evolved state $\left[  \exp\left[  i\phi\left(  a^{\dagger}a+b^{\dagger
}b\right)  \right]  \left\vert \psi(\tau)\right\rangle _{ab}\left\vert
i\right\rangle +\left\vert \psi(\tau)\right\rangle _{ab}\left\vert
f\right\rangle \right]  /\sqrt{2}$ where $\phi=\left\vert \lambda\right\vert
^{2}t/\Delta$ ($\left\vert \lambda\right\vert =\left\vert \lambda
_{a}\right\vert \sim\left\vert \lambda_{b}\right\vert $). After undergoing a
$\pi/2$ pulse in a Ramsey zone, with phase chosen so that $\left\vert
i\right\rangle \rightarrow\left(  \left\vert i\right\rangle +\left\vert
f\right\rangle \right)  /\sqrt{2}$ and $\left\vert f\right\rangle
\rightarrow\left(  \left\vert i\right\rangle -\left\vert f\right\rangle
\right)  /\sqrt{2}$, the atomic states $\left\vert i\right\rangle $ and
$\left\vert f\right\rangle $ are measured with probabilities $\mathcal{P}_{i}$
and $\mathcal{P}_{f}$. Finally, the direct measurement of the two-mode Wigner
function\ follows from $W(\eta_{a},\eta_{b},\eta_{a}^{\ast},\eta_{b}^{\ast
})\varpropto\mathcal{P}_{f}-\mathcal{P}_{i}$ \cite{FLD}. In the particular
case of degenerated parametric down-convention, where the resulting
Hamiltonian is the squeezing operator of a single cavity mode, the same scheme
can be used to measure directly the Wigner function of any squeezed state.

\begin{acknowledgments}
We wish to express thanks for the support of FAPESP (under contracts
\#99/11617-0, \#00/15084-5, and \#02/02633-6) and CNPq (Instituto do
Mil\^{e}nio de Informa\c{c}\~{a}o Qu\^{a}ntica), Brazilian research funding agencies.
\end{acknowledgments}


\begin{thebibliography}{99}                                                                                               %


\bibitem {Stoler}D. Stoler, Phys. Rev. Lett. \textbf{33}, 1397 (1974).

\bibitem {Kwiat}P. G. Kwiat, \textit{et al.}, Phys. Rev. Lett. \textbf{75},
4337 (1995).

\bibitem {Louisell}W. H. Louisell, A. Yariv, and A. E. Siegman, Phys. Rev.
\textbf{124}, 1646 (1961).

\bibitem {Milburn}G. J. Milburn, Opt. Acta. \textbf{31}, 671 (1984); H. J.
Kimble, M. Dagenais, and L. Mandel, Phys. Rev. Lett. \textbf{39,} 691 (1977);
H. P. Yuen and J. H. Shapiro, Opt. Lett. \textbf{4}, 334 (1979).

\bibitem {Haroche}M. Brune, \textit{et al}., Phys. Rev. Lett. \textbf{76},
1800 (1996); D. M. Meekhof, \textit{et al}., \textit{ibid}. \textbf{76}, 1796
(1996); Ch. Roos, \textit{et al}.,\textit{\ ibid}. \textbf{83}, 4713 (1999).

\bibitem {SN}H. P. Yuen and J. H. Shapiro, IEEE Trans. Inf. Theory
\textbf{24}, 657 (1978); D. J. Wineland, J. J. Bollinger, W. M. Itano, and D.
J. Heinzen, Phys. Rev. A \textbf{50}, 67 (1994).

\bibitem {Caves}J. N. Hollenhorst, Phys. Rev. D \textbf{19}, 1669 (1979); C.
M. Caves, \textit{et al}., Rev. Mod. Phys. \textbf{52}, 341 (1980).

\bibitem {Pittman}E. Knill, R. Laflamme, and G. J. Milburn, Nature
\textbf{409}, 46 (2001); T. B. Pittman, B. C. Jacobs, and J. D. Franson, Phys.
Rev. Lett. \textbf{88}, 257902 (2002).

\bibitem {CZ}J. I. Cirac and P. Zoller, Phys. Rev. Lett. \textbf{74}, 4091
(1995); T. Pellizzari, \textit{ibid}. \textbf{79}, 5242 (1997); S. Lloyd, M.
S. Shahriar, J. H. Shapiro, and P. R. Hemmer, \textit{ibid}. \textbf{87},
167903 (2001).

\bibitem {Rausch}A. Rauschenbeutel, \textit{et al}., Phys. Rev. A \textbf{64},
050301(R) (2001); B. T. H. Varcoe, S. Brattke, M. Weidinger, H. Walther,
Nature \textbf{403,} 743 (2000).

\bibitem {EPR}A. Einstein, B. Podolsky, and N. Rosen, Phys. Rev. \textbf{47},
777 (1935).

\bibitem {Walls}C. A. Blockley and D. F. Walls, Phys. Rev. A, \textbf{43},
5049 (1991); N. A. Ansari, J. Gea--Banacloche, and M. S. Zubairy,
\textit{ibid}. \textbf{41}, 5179 (1990); B. J. Dalton, Z. Ficek, and P. L.
Knight, \textit{ibid}. \textbf{50}, 2646 (1994).

\bibitem {Gerry}C. C. Gerry and J. H. Eberly, Phys. Rev. A \textbf{42}, 6805 (1990).

\bibitem {Duvida}S. C. Gou, Phys. Rev. A \textbf{40}, 5116 (1989).

\bibitem {Celso}C. J. Villas-B\^{o}as, N. G. de Almeida, R. M. Serra, and M.
H. Y. Moussa, Phys. Rev. A \textbf{68}, 061801(R) (2003).

\bibitem {SB}S. L. Braunstein,C. A. Fuchs, H. J. Kimble, and P. van Loock ,
Phys. Rev. A \textbf{64}, 022321 (2001).

\bibitem {BRH}M. Brune, J. M. Raimond, and S. Haroche, Phys. Rev. A
\textbf{35}, 154 (1987).

\bibitem {Allen}For the typical parameters mentioned here, with $\Delta
\sim14\lambda_{a,b}$, we have a deviation from unity about 3\%, for
$\left\langle \sigma_{ii}\right\rangle (t)$, when considering a numerical
simulation with Hamiltonian (\ref{Eq2}). A carefull analyses, including
numerical simulations, will be present elsewhere. A similar discussion has
been done in Ref. X. X. Yi, \textit{et al}, e-print quant-ph/0306035.

\bibitem {Furusawa}A. Furasawa, \textit{et al}., Science \textbf{282}, 706
(1998); E. S. Polzik, J. Carri, and H. J. Kimble, Phys. Rev. Lett.
\textbf{68}, 3020 (1992).

\bibitem {Haroche01}S. Osnaghi, \textit{et al.}, Phys. Rev Lett. \textbf{87},
037902 (2001).

\bibitem {Celso01}C. J. Villas-B\^{o}as, F. R. de Paula, R. M. Serra, and M.
H. Y. Moussa, Phys. Rev. A \textbf{68}, 053808 (2003).

\bibitem {FLD}M. Fran\c{c}a Santos, L. G. Lutterbach, and L.
Davidovich\textit{,} J. Opt. B: Quantum Semiclass. Opt. \textbf{3}, S55
(2001); L. G. Lutterbach and L. Davidovich, Phys. Rev. Lett. \textbf{78}, 2547 (1997).
\end{thebibliography}
\end{document}